# On Wavelet Decomposition over Finite Fields

H.M. de Oliveira, T.H. Falk

CODEC – Communications Research Group
Departamento de Eletrônica e Sistemas – CTG – UFPE
C.P. 7800, 50711-970, Recife – PE, Brazil
e-mail: hmo@ufpe.br, tiagofalk@go.com

**Abstract**–*This paper introduces some foundations of wavelets over Galois fields. Standard orthogonal finite-field wavelets (FF-Wavelets) including FF-Haar and FF-Daubechies are derived. Non-orthogonal FF-wavelets such as B-spline over GF(p) are also considered. A few examples of multiresolution analysis over Finite fields are presented showing how to perform Laplacian pyramid filtering of finite block lengths sequences. An application of FF-wavelets to design spread-spectrum sequences is presented.*

## 1. INTRODUCTION

Continuos and Discrete Wavelet transforms have emerged as a powerful tool in signal analyses and have proven to be superior to classical Fourier analysis in many situations [REI 94, POLI 00]. There has been a proliferation of wavelet applications including seismic geology, quantum physics, medical area, image processing (e.g., video data compression, reconstruction of high resolution images), computer graphics, filter banks, and so on. Essentially, the Wavelet Transform is a signal decomposition onto a set of basis function, which are derived from a single prototype wavelet by scaling (dilatations and contractions) as well as translations (shifts) [GOM et al. 97].

Fourier analysis can also be carried out over finite fields. The most known of such tools is the Finite Field Fourier transform, introduced by Pollard [POL 71]. Another finite field transform was recently introduced [CAM et al. 98]. Those kinds of transforms play an important role in problems related with the finite field structure. It seems to be quite natural to think about a possible wavelet analysis over a finite field, which could present some advantages regarding the "classical" finite field Fourier analysis.

The basic FF-wavelet is an N-dimensional vector $\underline{\psi}_{1,0} = (\psi_{1,0}(0), \psi_{1,0}(1), \psi_{1,0}(2),..., \psi_{1,0}(N-1))$, where each component of $\psi$ belongs to the extension field $GF(p^s)$. We begin with FF-wavelets over $GF(p)$. Let N be an integer and $D(N)$ be the set of the divisors of N. On finite fields scaling cannot be carried out by a real number, $a \in \Re$, as usual but on divisor of the length. The following operations are allowed:

1) scaling $\underline{\psi}_{j,0}$ where $\psi_{j,0}(i) = \psi_{1,0}(ji)$,

$\forall\, j \in D(N/2) := \{\,j \text{ such that } j \mid N/2\,\}$.

2) translation $\underline{\psi}_{j,k}$ where

$\psi_{j,k}(i) = \psi_{j,0}(i + \dfrac{Nk\,(\text{mod } N)}{j})$, $\forall$ k=0,1,...N-1.

The wavelet functions are
$\underline{\psi}_{j,k} = (\psi_{j,k}(0),\ \psi_{j,k}(1),\ \psi_{j,k}(2),...,\ \psi_{j,k}(N-1))$,
which are scaled and/or translated versions of the basic FF-wavelet $\underline{\psi}_{1,0}$.

**Property 1**: $\sum_{i=0}^{N-1} \psi_{j,k}(i) \equiv 0\,(\text{mod } p)$ $(\forall j,k)$.

Let $\underline{v}=(v_0, v_1,..., v_{N-1})$ be a signal-vector of blocklength N over a Galois field GF(p), of characteristic $p \neq 2$ and $\underline{\psi}_{j,k}$ wavelets functions over $GF(p^s)$, $s \geq 1$.

**Notation.** The FF-wavelet Transform of a signal $\underline{v}$ is defined by $FFWT(j,k) \equiv \sum_{i=0}^{N-1} v_i \psi_{j,k}(i)\,(\text{mod } p)$, which is denoted by $FFWT(j,k) = <\underline{v}, \underline{\psi}_{j,k}>$. ∎

## 2. HAAR DECOMPOSITION OVER FINITE FIELDS

In this section we present the design of generalised Haar orthogonal bases. Assume that $p \equiv \pm 1\,(\text{mod } 8)$ and that N is a power of two.

**Definition 1.** A basic FF-Haar wavelet is defined according to
$$\psi_{1,0}(i) = \begin{cases} 1 & \text{if } 0 \leq \dfrac{i}{N} < \dfrac{1}{2} \\ p-1 & \text{if } \dfrac{1}{2} \leq \dfrac{i}{N} < 1 \\ 0 & \text{otherwise} \end{cases}$$

i.e., $\psi_{1,0}(i) = \begin{cases} (p-1)^{\left\lfloor \frac{i \pmod N}{N/2} \right\rfloor} & \text{if } 0 \leq \frac{i}{N} < 1 \\ 0 & \text{otherwise} \end{cases}$ ∎

Example 1: Let us consider the FF-Haar, N=8, over GF(p). Possible scaling factors $j \in D(4) = \{1, 2, 4\}$. Therefore

j=1  $\underline{\psi}_{1,0} = (1, 1, 1, 1, p-1, p-1, p-1, p-1)$
j=2  $\underline{\psi}_{2,0} = (1, 1, p-1, p-1, 0, 0, 0, 0)$
j=4  $\underline{\psi}_{4,0} = (1, p-1, 0, 0, 0, 0, 0, 0)$.

Translations of such sequences, e.g.,
$\underline{\psi}_{2,1} = (0, 0, 0, 0, 1, 1, p-1, p-1)$, are allowed.

**Property 2**. Given a scaled version $\underline{\psi}_{j,0}$, the number of different translated versions of a wavelet $\underline{\psi}_{j,k}$ is equal to j.

Energy normalisation. Since N is a power of two and j belongs to D(N/2), N/j is also a power of two. Supposing now that $p \equiv \pm 1 \pmod 8$, then $\sqrt{\frac{N}{j}} \in$ GF(p). Therefore, normalised transforms can be defined by

$$FFWT(j,k) = \frac{1}{\sqrt{\left(\frac{N}{j}\right) \pmod p}} \sum_{i=0}^{N-1} v_i \psi_{j,k}(i) \pmod p.$$

We consider a subset $S \subseteq D(N/2)$ such that $\sum_{j \in S \subseteq D(N/2)} j = N-1$.

When $N = 2^m$, then $D(N/2) = \{1, 2, 4, 8, \ldots 2^{m-1}\}$ and $\sum_{j \in D(N/2)} j = \sum_{j=1}^{m-1} 2^j = N-1$ so that all values of $j \in D(N/2)$ are used to derive scaled versions of the basic wavelet. These waveforms together with the signal ( 1 1 1 1 ... 1 ) generate N-orthogonal signals of blocklength N over GF(p), i.e., it generates an orthogonal Haar bases.

Example 2. GF(7)-Haar and normalised FF-Haar wavelet over GF(7).

( 1 1 1 1 1 1 1 1 )     ( 1 1 1 1 1 1 1 1 )
( 1 1 1 1 6 6 6 6 )     ( 6 6 6 6 1 1 1 1 )
( 1 1 6 6 0 0 0 0 )     ( 4 4 3 3 0 0 0 0 )
( 0 0 0 0 1 1 6 6 )     ( 0 0 0 0 4 4 3 3 )
( 1 6 0 0 0 0 0 0 )     ( 5 2 0 0 0 0 0 0 )
( 0 0 1 6 0 0 0 0 )     ( 0 0 5 2 0 0 0 0 )
( 0 0 0 0 1 6 0 0 )     ( 0 0 0 0 5 2 0 0 )
( 0 0 0 0 0 0 1 6 )     ( 0 0 0 0 0 0 5 2 ).

Clearly,
$\sum_{i=0}^{N-1} \psi_{j,k}(i) \equiv 0 \pmod p$,
$\sum_{i=0}^{N-1} \psi^2_{j,k}(i) \equiv 1 \pmod p$ and
$\sum_{i=0}^{N-1} \psi_{j,k}(i) \psi_{j',k'}(i) \equiv 0 \pmod p$ $\forall j \neq j'$ or $k \neq k'$.

If N is not a power of two, non-orthogonal wavelets can be derived, e.g., N=24 over GF(7). In this case, D(12)={1, 2, 3, 4, 6, 12}.

|  | # of translated versions |
|---|---|
| 1 1 1 1 1 1 1 1 1 1 1 1 1 1 1 1 1 1 1 1 1 1 1 1 | 1 |
| j=1 | |
| 1 1 1 1 1 1 1 1 1 1 1 1 6 6 6 6 6 6 6 6 6 6 6 6 | 1 |
| j=4 | |
| 1 1 1 6 6 6 0 0 0 0 0 0 0 0 0 0 0 0 0 0 0 0 0 0 | 4 |
| j=6 | |
| 1 1 6 6 0 0 0 0 0 0 0 0 0 0 0 0 0 0 0 0 0 0 0 0 | 6 |
| j=12 | |
| 1 6 0 0 0 0 0 0 0 0 0 0 0 0 0 0 0 0 0 0 0 0 0 0 | 12 |
|  | 24 |

These wavelets are no more orthogonal, for instance, $<\underline{\psi}(2,1), \underline{\psi}(6,0)> \neq 0 \pmod p$. The dual wavelets can easily be derived.

### 3. HAAR PYRAMID DECOMPOSITION

One of the most powerful tools in wavelet theory is the decomposition of data by the pyramid-filtering algorithm [COH 90, REI 94, GOM et al. 97].

Example 3. The two-element filter (1, 1) for the FF-Haar decomposition over GF(7) is,

h= [ 5  5 ]     h* = [ 5  5 ]
g= [ 2  5 ]     g* = [ 5  2 ]              ∎

### 4. B-SPLINE OVER FINITE FIELD

A simple non-orthogonal wavelet such as $n < p+1$ cardinal B-spline can be easily derived over GF(p), where $p \equiv \pm 1 \pmod 8$. An (n+2)-element filter is given by $(2^n)^{-1} \binom{n+1}{k} \pmod p$, k=0,1,2,...,n+1. The normalising factor is $\sqrt{2} \pmod p \cdot 2^{-1}$. Therefore, a quadratic spline ( 2-cardinal B-spline) over GF(7) is given by [ 3 2 2 3 ] or [ 4 5 5 4].

### 5. FF-DAUBECHIES WAVELETS OVER FINITE PRIME FIELD

Perhaps the most important orthogonal wavelet is Daubechies wavelet [DAU 88]. We want to find

the standard orthogonal wavelet decomposition. We deal with wavelets derived from a prototype $\psi(.)$:

$$\psi_{j,k}(i) = (\sqrt{2})^j \psi(2^j i - k).$$

The multiresolution analysis is generated by a scaling function $\phi$ such that:

$\phi_{j,k}(i) \equiv (2)^{j/2} \phi(2^j i - k)$

so that $\phi_{1,k}(i) \equiv \sqrt{2} \cdot \phi(2i - k)$ (mod p).

An orthogonal multiresolution over GF(p) is generated by the scaling function $\phi$ that holds the

Dilatation or Refinement Equation.

$$\phi(i) \equiv \sqrt{2} \pmod{p} \sum_{k=0}^{N-1} h_k \phi(2i - k) \pmod{p}.$$

The FF-wavelet must satisfy:

$$\psi(i) \equiv \sqrt{2} \pmod{p} \sum_{k=0}^{N-1} g_k \phi(2i - k) \pmod{p}.$$

Thus, the filter h and g are mirror en quadrature and a multiresolution of length $N=2^n$ is performed in n steps.

Requirements of filters for multiresolution.

$$\sum_{k=0}^{N-1} h_k \equiv \sqrt{2} \pmod{p}, \quad \sum_{k=0}^{N-1} g_k \equiv 0 \pmod{p}.$$

$$g_k \equiv (-1)^k h_{N-1-k} \pmod{p},$$

and $\sum_{k=0}^{N-1} h_k h_{k+2m} \equiv 0 \pmod{p}$, m≠0. ∎

Example 5. An N=4 FF-Daubechies over GF(97). The following filters perform the multiresolution analysis:
Smoothing (low-pass)  h = [ 92, 47, 12, 57]
Detail    (high-pass) g = [ 57, 85, 47, 5 ].
The filter are such that $g_k \equiv (-1)^k h_{3-k} \pmod{97}$.

Furthermore, $\sum_{k=0}^{3} h_k \equiv 14 \pmod{97}$,

$\sum_{k=0}^{3} g_k \equiv 0 \pmod{97}$ and $\sum_{k=0}^{3} h_k h_{k+2} \equiv 0 \pmod{97}$.
∎

## 6. APPLICATION: DESIGN OF SPREAD-SPECTRUM SEQUENCES

Digital multiplex usually refers to Time Division Multiplex (TDM). However, it can also be achieved by Coding Division Multiplex (CDM), which has recently been the focus of interest, especially after the IS-95 standardisation [QUAL 92]. The CDMA is now becoming a popular multiple access schemes. In this section we introduce a new class of CDM/CDMA schemes based on finite-field wavelet transforms.

The wavelet digital carriers have the same duration T of an input modulation symbol, so that it carries N chips per data symbol. The interval of each wavelet-symbol is T/N and therefore the bandwidth expansion factor when multiplexing N channels may be roughly N, the same result as FDM and TDM/PAM. A naive and illustrative example is presented in the sequel by considering the design of (orthogonal) spread-spectrum sequences of length N=8, based on FF-Haar wavelets over GF(7). Other FF-wavelets can also be adopted over other fields.

Example 6. A FF-Haar Spread-Spectrum over GF(7).
Orthogonal wavelets (e.g. example 3) can be adopted as spread-spectrum sequences. Each user has a spread-spectrum, which corresponds to a scaled/translated version of the same basic wavelet. A scheme of such a system is shown in figure 2.

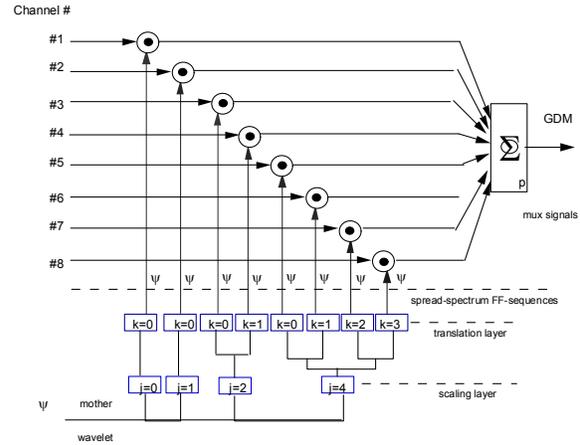

Figure 1. A Multiplex Based on FF-Haar Wavelets.

The operator $\Sigma$ denotes the conventional vectorial addition but with components taken modulo p. The duration of an information symbol at the input is N times greater than one GF(p)-symbol of the spread sequences. Thus, if data from the channel #3 is 2, $2 \in GF(7)$, and the spread-sequence of this channel is $\underline{\psi}_{2,0}$ =( 4 4 3 3 0 0 0 0 ), then the spread-signal is ( 2 2 2 2 2 2 2 2 ) $\otimes$ ( 4 4 3 3 0 0 0 0 ) $\equiv$ ( 1 1 6 6 0 0 0 0 ) (mod 7). For the sake of simplicity, this is denoted by $2^{(8)} \otimes 4^{(2)} 3^{(2)} 0^{(4)}$ ∫ $1^{(2)} 6^{(2)} 0^{(4)}$ (mod 7). Suppose for instance that, at some time, data from channels 1 to 8 are

$( 3\ 0\ 2\ 1\ 6\ 5\ 5\ 4\ )^T$ and the corresponding spreading-sequence (user's signature code) are

$(\underline{\psi}_{0,0}\ \underline{\psi}_{1,0}\ \underline{\psi}_{2,0}\ \underline{\psi}_{2,1}\ \underline{\psi}_{4,0}\ \underline{\psi}_{4,1}\ \underline{\psi}_{4,2}\ \underline{\psi}_{4,3})$.

The CDMed sequence will be:
$CDMed \equiv 3^{(8)} \oplus 0^{(8)} \oplus 1^{(2)}6^{(2)}0^{(4)} \oplus 0^{(4)}4^{(2)}3^{(2)}$
$\oplus 2^{(1)}5^{(1)}0^{(6)} \oplus 0^{(2)}4^{(1)}3^{(1)}0^{(4)} \oplus 0^{(4)}4^{(1)}3^{(1)}0^{(2)} \oplus 0^{(6)}6^{(1)}1^{(1)}$,

that is, CDMed$\equiv ( 6\ 2\ 6\ 5\ 4\ 3\ 5\ 0 )$ (mod 7):=$\underline{r}$. Clearly, the signal is neither TDMed nor FDMed. The multiplex and multiple access systems derived from the application of FF-wavelets can be viewed as a Galois-Field Multiple Access (GDMA) technique, recently introduced by the authors [deO et al. 99].

Since FF-Haar wavelets are orthogonal, data from each user can be easily retrieved by an inner product over GF(p):

channel #3  $<\underline{r}, \underline{\psi}_{2,0}> \equiv 2$ (mod 7), channel #8

$<\underline{r}, \underline{\psi}_{4,3}> \equiv 4$ (mod 7), etc.

The GDM system must guarantee a perfect synchronisation between casoidal carriers used at the mux and demux [deO-CAM 00, deO et al. 01]. Orthogonal-FFWT can be used as spread-spectrum sequences so as to implement new Galois-Division- synchronism control since all the user's sequences are derived from the same basic wavelet. Therefore, all the spread-sequences are generated from the same "clock", by scaling and shifts. Another attractive idea is to apply multiresolution to implement demultiplex. Although the implementation showed in the above example elucidates the multiplex mechanism, other wavelet issues such as multiresolution analysis can be used for (de)multiplexing.

## 7. CONCLUSIONS

The aims of this paper are to present new Finite Field techniques and show their potential applications. Finite Field Wavelets can be used as a powerful tool in the design of multilevel spread-spectrum sequences. Users have different categories of spread depending on the scaling factor. New digital multiplex schemes based on such sequences have also been introduced, which are multilevel Code Division Multiplex. This approach exploits orthogonality properties of synchronous non-binary sequences defined over a finite field.

### ACKNOWLEDGEMENTS
The authors wish to thank R.F.G. Távora for insightful comments, which improved the final version of this paper.